\documentstyle[prd,aps,preprint,tighten,epsfig]{revtex}
\begin{document}
\draft
\title{THE ADLER FUNCTION FOR LIGHT QUARKS IN ANALYTIC PERTURBATION THEORY}
\author{K.A. Milton\thanks{E-mail: milton@mail.nhn.ou.edu}}
\address{Department of Physics and Astronomy,
University of Oklahoma, Norman, OK 73019 USA}
\author{I.L. Solovtsov\thanks{E-mail: solovtso@thsun1.jinr.ru} and
O.P. Solovtsova\thanks{E-mail: olsol@thsun1.jinr.ru}}
\address{Bogoliubov Laboratory of Theoretical Physics,
 Joint Institute for Nuclear Research, 141980 Dubna, Moscow Region, Russia }
\preprint{OKHEP--01--02}
\date{\today} \maketitle
\def\si{\sigma}   \def\mm{M_{\tau}^2}

\begin{abstract}
The method of analytic perturbation theory, which avoids the problem of
ghost-pole type singularities and gives a self-consistent description of
both spacelike and timelike regions, is applied to describe the ``light"
Adler function corresponding to the non-strange vector channel of the
inclusive decay of the $\tau$ lepton. The role of threshold effects is
investigated. The behavior of the quark-antiquark system near threshold
is described by using a new relativistic resummation factor. It is shown
that the method proposed leads to good agreement with the ``experimental''
Adler function down to the lowest energy scale.
\end{abstract}


\section{Introduction}
In studying the relationship between theoretical predictions and experimental
data, it is important to connect measured quantities with the most elementary
theoretical objects to check direct consequences of the theory without
essential
use of model assumptions. Some single-argument functions
which are directly connected with experimentally measured quantities can
play the role of these objects. A theoretical description of inclusive
processes can be made in terms of functions of this sort. Among them
is the Adler function \cite{Adler} which can be extracted from the
experimental data for the process of $e^+e^-$ annihilation into hadrons and
the inclusive decay of the $\tau$ lepton. The mass of the $\tau$ lepton,
$M_\tau=1.777\,{\rm{GeV}}\,,$ is large enough in order to produce decays
with a hadronic mode. At the same time, in the context of QCD, the mass is
sufficiently small to allow one to investigate effects lying beyond the
framework of the perturbative approach. At present, there is rich
experimental material obtained from hadronic decays of the $\tau$ lepton.
The first
theoretical analysis of hadronic decays of a heavy lepton was performed in
1971~\cite{Tsai} well before the experimental discovery of the
$\tau$-lepton in
1975. Since then, the properties of the $\tau$ have been studied very
intensively.

The ratio of hadronic to leptonic widths for the inclusive decay of the
$\tau$-lepton,
$$R_{\tau}=\frac{\Gamma(\tau^{-}\to{\rm{hadrons}}\;\nu_{\tau})}
{\Gamma(\tau^{-}\to{\ell}\;{\bar{\nu}}_{\ell}\;{\nu}_{\tau})}\,,$$ is the
most precise one for extracting of the values of the fundamental QCD
parameters at a low energy scale \cite{PDG98}. The initial theoretical
expression for $R_{\tau}$ contains an integral over timelike
momentum
\begin{equation}\label{Eq.1}
R_{\tau} =  \frac{2}{\pi}\;  \int^{M_\tau^2}_0 {ds \over M_\tau^2 } \,
\left(1-{s \over M_\tau^2}\right)^2
\left(1 + 2 {s \over M_\tau^2}\right)
{\rm Im}\; \Pi(s) \;,
\end{equation}
which extends down to small $s$ and cannot be directly calculated in the
framework of standard perturbation theory~(PT). Indeed, the hadronic
correlation function
$\Pi(s)$ is parametrized by the perturbative running coupling
which has unphysical singularities and, therefore, is ill-defined in the
region of small momenta. To avoid this problem, one usually applies the
following procedure. The initial integral~(\ref{Eq.1}) is rewritten by using
the Cauchy theorem in the form of a contour integral in the complex plane
with the contour running around a circle with radius
$M_{\tau}^2$~\cite{BNP92,LP92}:
\begin{equation}\label{Eq.2}
R_{\tau}\, =\,\frac{1}{2\pi {\rm i}} \, \oint_{|z|=M_\tau^2}
\frac{d\,z}{z}\, \left(1-{z \over M_\tau^2}\right)^3 \left(1 + {z \over
M_\tau^2}\right) D(z) \, ,
\end{equation}
where $D(z)=-z \;\displaystyle{d\Pi(z)/dz}$
is the Adler function.
This trick allows one, in principle, to
avoid the problem of a direct calculation of the $R_{\tau}$ ratio by
Eq.~(\ref{Eq.1}). However, in order to perform this transformation
self-consistently, it is necessary to maintain correct analytic properties
of the hadronic correlation function,
which are violated in the framework of standard
PT. The analytic approach to QCD~\cite{SS96-97}, the so-called analytic
perturbation theory~(APT)~\cite{MSS97,Vanc98}, maintains needed analytic
properties
and allows one to give meaning to the initial expression. The APT
description can be equivalently phrased either on the basis of the
expression~(\ref{Eq.1}) or on the contour
representation~(\ref{Eq.2})~\cite{MSS97}.
The information obtained in $\tau$ measurements allows one to
construct various ``experimental'' curves. In particular, in paper
\cite{GNeubert95} a quantity $R_\tau(s_0)$ with a variable ``mass" $s_0\leq
M_\tau^2$, has been considered. This quantity, defined for timelike
momenta, is similar to the ``smeared" functions constructed, for example,
according to the Poggio-Quinn-Weinberg method \cite{PQW}. The $\tau$-decay
data allow us also to determine quantities in the Euclidean region,
including the Adler function. 
This function can be extracted from the $\tau$ data down to
the lowest energy scale \cite{PPRaf}.

In this paper we study the  Adler $D$-function corresponding to
$\tau$ decay mediated by the non-strange vector current via $W^-\to
d{\bar{u}}$. We use the APT method which does not encounter the problem of
unphysical singularities of the running coupling and gives a self-consistent
description of both the timelike, Eq.~(\ref{Eq.1}), and the spacelike,
Eq.~(\ref{Eq.2}), regions.

The region of integration in Eq.~(\ref{Eq.1}) includes the vicinity of the
quark-antiquark threshold. The perturbative expansion breaks down in this
neighborhood due to singularities at $s=(m_q+m_{\bar{q}})^2$
\cite{App-Politzer75,PQW}. Thus any finite order of the perturbative expansion
is unreliable near quark thresholds and, therefore, all singular terms of the
$(\alpha_S/v)^n$ type, where $v$ is the relative velocity of
the quarks, have to be summed. Note that this problem cannot be
avoided by using the contour representation
(\ref{Eq.2}) instead of Eq.~(\ref{Eq.1}),
because these expressions should be equivalent to each other
in the framework of a consistent method. For heavy quark systems one usually
uses the nonrelativistic resummation factor obtained by using the
Schr\"odinger equation with the Coulomb potential, which is known as the
Sommerfeld-Sakharov factor~\cite{Sommerfeld,Sakharov}. But for a systematic
description of the threshold region in the system of light quarks it is
necessary to apply a relativistic approach. Here, we take into account
threshold effects by using a new relativistic resummation factor proposed in
Ref.~\cite{MS-Sfac00}, which was obtained for a QCD-like potential.

\section{Analytic approach to $\tau$ decay}
We start our consideration with a three-loop PT and APT analysis,
neglecting, in the beginning, quark masses. It is convenient to separate
the QCD contribution by representing the $R_{\tau}$ ratio in the form
\begin{equation}
R_{\tau}=R_{\tau}^{0}(1+\delta_{\rm QCD}),
\end{equation}
where $R_{\tau}^{0}$ corresponds to the parton level description and
$\delta_{\rm QCD}$ is the QCD correction. We introduce QCD contributions to
the
imaginary part of the hadronic correlator, $r(s)$, and to the corresponding
Adler function, $d(z)$ as follows:
$${\cal{R}}(s)={\rm{Im}}\;\Pi(s+i\epsilon)/\pi R_\tau^0 \propto1+r \, ,\,
\qquad D(z)\propto1+d(z) \,.$$
Then, one can write $\delta_{\rm QCD}$ as an integral over timelike momentum
(Minkowskian region)
\begin{equation}\label{to-tau}
\delta_{\rm QCD}=2\int_0^{\mm}\frac{ds}{\mm}{\left(1-\frac{s}{\mm}
\right)}^2\left(1+2\frac{s}{\mm}\right) r(s) \, ,
\end{equation}
or as a contour integral in the complex plane (Euclidean region)
\begin{equation} \label{contour}
\delta_{\rm QCD}=\frac{1}{2\pi
i}\oint_{|z|=\mm}\frac{dz}{z}{\left(1-\frac{z}{\mm}\right)}^3
\left(1+\frac{z}{\mm}\right) d(z)\, .
\end{equation}

We now consider $\delta_{\rm QCD}$ in the framework  the PT and APT methods.

\subsection{Perturbation theory}
The PT description is based on the contour representation and can be
developed in the following two ways. In Braaten's (Br) method \cite{BNP92}
the quantity (\ref{contour}) is represented in the form of truncated power
series with the expansion parameter
$a_\tau\equiv{\alpha_S(M_\tau^2)}/{\pi}$. In this case the three-loop
representation for $\delta_{\rm QCD}$ is
\begin{equation} \label{delta_Br}
\delta_{\rm QCD}^{\rm Br} = a_\tau \, + r_1\, a_\tau^2 + r_2\,a_\tau^3 \> ,
\end{equation}
where the coefficients $r_1$ and $r_2$ in the $\overline{\rm{MS}}$ scheme
with three active flavors are $r_1=5.2023$ and $r_2=26.366$~\cite{BNP92}.

The method proposed by Le~Diberder and Pich (LP) \cite{LP92} uses the PT
expansion of the $d$-function
\begin{equation}\label{d_expand}
d(z)=a(z)+ d_1\, a^2(z)+d_2\, a^3(z) \, ,
\end{equation}
where in the $\overline{\rm{MS}}$-scheme $d_1=1.640$ and
$d_2=6.371$~\cite{r35} for three active quarks.
The three-loop PT running coupling, $a(z)$, is commonly written in the form of
an expansion in inverse powers of $L$~\cite{PDG98}.
In the $\overline{\rm{MS}}$ scheme it is
\begin{equation}\label{PT3}
a(z)=\frac{4}{\beta_0 L}\left\{ 1- \frac{\beta_1}{\beta_0^2}\frac{\ln L}{L}
+ \frac{1}{L^2} \left[\, \frac{\beta_1^2}{\beta_0^4}
\left(\ln^2L-\ln\,L-1\right) +  \frac{\beta_2}{\beta_0^3} \right ] \right\}
\,, \>\> L \equiv \ln (-z/\Lambda^2) \, ,
\end{equation}
where $\beta_0=11-2n_f/3$, $\beta_1=102-38n_f/3$ and
$\beta_2^{\overline{\rm{MS}}}=2857/2-5033n_f/18+325n_f^2/54$
are the first three $\beta$-function coefficients.
The substitution of Eq.~(\ref{d_expand}) into Eq.~(\ref{contour}) leads to the
following non-power representation
\begin{equation}\label{delta_LP}
\delta_{\rm QCD}^{\rm LP} = A^{(1)}(a)+ d_1 \,A^{(2)}(a)+d_2\,A^{(2)}(a)
\end{equation}
with
\begin{equation}\label{A}
A^{(n)}(a)=\frac{1}{2\pi
i}\oint_{|z|=\mm}\frac{dz}{z}{\left(1-\frac{z}{\mm}\right)}^3
\left(1+\frac{z}{\mm}\right) a^{n}(z)\, .
\end{equation}

As noted above, transformation to the contour representation~(\ref{contour})
requires the existence of certain analytic properties of the correlator:
namely, it must be an analytic function in the complex $z$-plane with a cut
along the positive real axis. The correlator parametrized, as usual, by the
PT running coupling does not have this virtue. Moreover, the conventional
renormalization group method determines the running coupling in the
spacelike region, whereas the initial expression (\ref{Eq.1}) contains an
integration over timelike momentum, and there is the question of how to
parametrize a quantity defined for timelike momentum transfers~\cite{RKP82}.
To perform this procedure self-consistently, it is important to maintain
correct analytic properties of the hadronic correlator
\cite{s10,MSS-time,Shirkov-time}. Because of this failure of analyticity,
Eqs.~(\ref{to-tau}) and (\ref{contour}) are not equivalent in the framework
of PT and, if one remains within PT, it is difficult to estimate the errors
introduced by this transformation. However, using the APT method, it is
possible
to resolve these problems.\footnote{The nonperturbative $a$-expansion
technique
in QCD~\cite{s1} also leads to a well-defined procedure of analytic
continuation~\cite{s10}.}

\subsection{Analytic perturbation theory}
In the framework of the analytic approach,\footnote{To distinguish APT and
PT cases, we will use subscripts ``an'' and ``pt''.} the functions $d(z)$
and $r(s)$ are expressed in terms of the effective spectral function
$\rho(\si)$ \cite{SS96-97,MSS-time}
\begin{equation} \label{drho-rrho}
d(z)=\frac{1}{\pi}\int^{\infty}_0 \frac{d\si}{\si - z}\, \rho(\si)\,,
\qquad
r(s)=\frac{1}{\pi}\int^{\infty}_{s}\frac{d\si}{\si} \rho(\si)\, .
\end{equation}
The APT spectral function is defined as the imaginary part
of the perturbative approximation to $d_{\rm pt}$ on the physical cut
\begin{equation} \label{rho}
\rho (\si)=\varrho_0 (\si)+d_1 \varrho_1 (\si)+d_2\varrho_2 (\si)\,,
\end{equation}
where
\begin{equation} \label{rho_n}
\varrho_n (\si)={\rm Im}[a_{\rm pt}^{n+1}(\si+i\epsilon)]\,.
\end{equation}

The function $\varrho_0 (\si)$ in Eq.~(\ref{rho}) defines the analytic
spacelike, $a_{\rm an}(z)$, and timelike, $\tilde{a}_{\rm an}(s)$,
running couplings as follows
\begin{equation}\label{a_t-a_s}
a_{\rm an}(z)=\frac{1}{\pi}\int_0^\infty\frac{d\si}{\si-z}\,
\varrho_0(\si)\,, \qquad
\tilde{a}_{\rm an}(s)=
\frac{1}{\pi}\int^{\infty}_{s}\frac{d\si}{\si}\varrho_0(\si).
\end{equation}
As has been argued from general principles, the behavior of these couplings
cannot be the same~\cite{MS99}. It should be stressed that,
unlike the PT running
coupling, the analytic running coupling has no
unphysical singularities (the ghost pole and branch
points) and, therefore, possesses the correct analytic
properties, arising from  K\"all\'en-Lehmann analyticity reflecting the
general principles of the theory.
For example, the one-loop APT result is~\cite{SS96-97,MSS-time}
\begin{equation}
\label{a1}
a^{(1)}_{\rm an}(z)\;=\;a^{(1)}_{\rm pt}(z)\,+\,
\frac{4}{\beta_0}\frac{\Lambda^2}{\Lambda^2+z}\, , \> \> \>
\tilde{a}^{(1)}_{\rm an}(s)\,=\,\frac{4}{\beta_0}\left[
\frac{1}{2}-\frac{1}{\pi}\arctan \frac{\ln( s/\Lambda^2)}{\pi}
\right] \,,
\end{equation}
where $a^{(1)}_{\rm pt}(z)\;=\;4/\left[\beta_0 \ln(-z/\Lambda^2)\right]\,$.

The analytic running couplings (the exact two-loop and the three-loop after
an approximation) can be written explicitly in the term of the Lambert
function \cite{APT-Lambert,Gardi}. However, in the framework of the APT
approach
there is a little sensitivity to the approximation in solving the
renormalization group equation for the running coupling \cite{MSS97,Vanc98}.
In the following, we use an explicit form for the analytic running coupling
in the timelike region, $\tilde{a}_{\rm an}(s)$, that is derived by using
the formula (\ref{PT3}) in  Eq.~(\ref{rho}). The corresponding expression
reads
\begin{equation}\label{time3}
\tilde{a}_{\rm an}(s)\,=\,\tilde{a}_{\rm an}^{(1)}(s)\,+\,
\tilde\Delta_1(s)\,+
\,\tilde\Delta_2(s) \,,
\end{equation}
where $\tilde{a}^{(1)}_{\rm an}(s)$ is given in Eq.~(\ref{a1}) and
\begin{eqnarray}\label{time_2_3}
&&\tilde\Delta_1\,=\,-
\frac{4\beta_1}{\beta_0^3}\frac{1}{B^2}
\left[{\ln B}+1-A \,\frac{\ln (s/\Lambda^2)}{\pi} \right] \,,
\\
&&\tilde\Delta_2=\frac{4\pi\beta_1^2}{\beta_0^5}\frac{1}{B^4}
\left\{ \left( \frac{\beta_2\beta_0}{\beta_1^2}-1 -A^2+\ln^2 B\right)
\frac{\ln (s/\Lambda^2) }{\pi}  + \left[1-\frac{\ln^2 (s/\Lambda^2)}{\pi^2}
\right]A\, \ln B \right\}\, ,\nonumber \\
&&\quad A(s)= \frac{\beta_0}{4}\pi \;\tilde{a}^{(1)}_{\rm an}(s)\, , \quad
B(s) \equiv\sqrt{\pi^2+\ln^2 (s/\Lambda^2)}\,. \nonumber
\end{eqnarray}
For the QCD correction $r(s)$ one can write down~\cite{GI99}
\begin{equation}\label{r3}
r_{\rm an}(s)\,=\,\tilde{a}_{\rm an}(s)\,+\,\Delta{r}_1(s)\,+
\,\Delta{r}_2(s) \,,
\end{equation}
where $\tilde{a}_{\rm an}(s)$ is given by Eq.~(\ref{time3}) and
\begin{eqnarray}\label{r_2_3}
&&\Delta{r}_1\,=\,d_1
\left(\frac{4}{\beta_0}\right)^2\frac{1}{B^2}
\left\{ 1-\frac{\beta_1}{\beta_0^2}\frac{\pi}{B^2}
\left[\left(2\ln B+1\right)
\frac{\ln (s/\Lambda^2)}{\pi}  +A-A\frac{\ln^2 (s/\Lambda^2)}{\pi^2}
\right]\right\}\, ,
\nonumber \\
&&\Delta{r}_2=d_2\left(\frac{4}{\beta_0}\right)^3\frac{1}{B^4}
\ln (s/\Lambda^2) \,.
\end{eqnarray}

Using Eq.~(\ref{to-tau}) or equivalently Eq.~(\ref{contour}), we obtain the
QCD correction to the $R_{\tau}$-ratio in terms of $\rho(\si)$ as follows
\begin{equation}\label{danrho}
\delta_{\rm an}=\frac{1}{\pi}\int_{M_{\tau}^2}^{\infty}\frac{d\si}{\si}\rho
(\si)\,+\,
\frac{1}{\pi}\int^{\mm}_0\frac{d\si}{\si}\left[ 2\frac{\si}{\mm}-
2\left(\frac{\si}{\mm}\right)^3 +
\left(\frac{\si}{\mm}\right)^4\right]\rho (\si) \, .
\end{equation}
It is obvious that the first term  of this expression is $r_{\rm an}$
evaluated at the $\tau$ mass
and that $\delta_{\rm an}$ is not representable as a series expansion
in the running coupling.

The difference between the PT~(LP) and APT contributions to  $R_\tau$ can
be transparently shown by the one-loop relation:
\begin{equation}\label{delta1}
\delta^{(1)}_{\rm an}= \delta^{(1)}_{\rm pt}-\frac{8}{\beta_0}
\frac{\Lambda^2}{M_{\tau}^2}+
{\rm O}({\Lambda^4}/{M_{\tau}^4}) \, .
\end{equation}
The additional term, which is ``invisible'' in the perturbative expansion,
turns out to be important numerically~\cite{MSS97,OlSol96}.
Due to the negative sign of this term,
the QCD scale is larger in this method, $\Lambda_{\rm an} > \Lambda_{\rm
pt}$, at the same value of the QCD correction:
$\delta_{\rm an}(\Lambda_{\rm an})=\delta_{\rm pt}(\Lambda_{\rm pt})=
\delta^{\rm exp}$.
It should be noted that due to the difference of shapes of the APT and PT
running couplings (see, for example, Ref.~\cite{GLS98}), their values
at the $\tau$ scale do not differ very much \cite{tau00}.

The APT analysis of the $\tau$ decay in the three-loop level has been
performed in Ref.~\cite{tau00}. This investigation together with other results
(see, for example, Refs.~\cite{Vanc98,SS9899,Sh-nonpower}) allows us to
formulate
the following features of the APT method: (i) this approach maintains the
correct analytic properties and leads to a self-consistent procedure of
analytic continuation from the spacelike to the timelike region; (ii) it has
much improved convergence properties and turns out to be stable with respect
to higher-loop corrections; (iii) renormalization scheme dependence of the
results obtained within this method is reduced dramatically.

\subsection{The vector channel in $\tau$ decay}
Experimentally the $R_{\tau}$-ratio can be separated into three
parts
\begin{equation}
R_{\tau} \,= \,R_{\tau,V}\,+\,R_{\tau,A}\,+\,R_{\tau,S} \, .
\end{equation}
The terms $R_{\tau,V}$ and  $R_{\tau,A}$ are contributions coming from the
non-strange hadronic decays associated with vector ($V$) and axial-vector
($A$) quark currents respectively, and $R_{\tau,S}$ contains strange
decays ($S$).

Within the perturbative approximation with massless quarks the vector
and axial-vector contributions to $R_\tau$ coincide with each other
\begin{equation}
R_{\tau,V}=R_{\tau,A}=\frac{3}{2}|V_{ud}|^2(1\,+\,\delta_{\rm QCD}) \,,
\end{equation}
where $|V_{ud}|$ denotes the Cabibbo--Kobayashi--Maskawa matrix element.
However, the experimental measurements~\cite{ALEPH98,OPAL99} show that
these components are not equal to each other. The corresponding difference
is associated with non-perturbative QCD effects which are usually described
in the form of power corrections. The experimental data for the isovector
spectral function of the ALEPH Collaboration~\cite{ALEPH98} have been used
in Ref.~\cite{PPRaf} to extract the Adler $D_{V}$-function which we show as
the
dashed line in Fig.~\ref{d-v}. The experimental $D$-function turns out to be
a smooth function without any trace of resonance structure. The
$D$-function obtained in Ref.~\cite{EJKV99} from the data for
electron-positron
annihilation into hadrons also has a similar property. One can expect
that the Adler $D$-function defined in the Euclidean region reflects more
adequately the quark-hadron duality than do quantities determined in
the Minkowskian region,\footnote{The Minkowskian and Euclidean
characteristics which are associated with the process of
electron-positron annihilation
into hadrons have been considered in Ref.~\cite{MSS00}.} and, therefore,
is more suitable for relating theoretical predictions with experimental
data. It is obvious that the PT approximation cannot be applied at the low
energy scale. Besides, any finite order of the operator product expansion
fails to describe the infrared tail of the $D$-function. APT's good behavior
in the infrared region, in principle, allows us to consider the
$D$-function down to a low energy scale.

\section{The vector channel $D$-function}
In this section we describe the Adler function corresponding to inclusive
$\tau$ decay. By using a dispersion relation for the hadronic correlator
$\Pi(q^2)$, one can represent the Adler function as follows
\begin{equation}\label{D_R}
D(Q^2)=Q^2\,\int_0^{\infty}{ds}\,\frac{{\cal R}(s)}{(s+Q^2)^2}\, .
\end{equation}

We will derive ${\cal R}(s)$ by applying the APT approach while incorporating
threshold resummation.

\subsection{Threshold effects}
A description of quark-antiquark systems near threshold requires us to take
into account the resummation factor which summarizes the threshold
singularities of the perturbative series of the $(\alpha_S/v)^n$ type. In a
nonrelativistic approximation, this is the well known Sommerfeld-Sakharov
factor~\cite{Sommerfeld,Sakharov}. For a systematic relativistic analysis of
quark-antiquark systems, it is essential from the very beginning to have a
relativistic generalization of this factor. Moreover, it is important to
take into account the difference between the Coulomb potential in the case of
QED and the quark-antiquark potential in the case of QCD. This QCD
relativistic factor has been proposed in Ref.~\cite{MS-Sfac00} to have the
form
\begin{equation}  \label{S-factor-relativistic}
S(\chi)=\frac{X(\chi)}{1-\exp\left[-X(\chi)\right]}\, ,
\quad\quad X(\chi)=\frac{4\pi\,\alpha_S}{3\sinh\chi}\, ,
\end{equation}
where $\chi$ is the rapidity which related to $s$ by $2m\cosh\chi=\sqrt{s}$.
The relativistic resummation factor (\ref{S-factor-relativistic}) reproduces
both the expected nonrelativistic and ultrarelativistic limits and
corresponds to a QCD-like quark-antiquark potential.

A convenient way to incorporate quark mass effects is to use an
approximate expression~\cite{PQW,App-Politzer75} which here can be written as
\begin{equation}\label{R_approx}
{\cal{R}}(s)= T(v)\, \left[ 1\,+\,g(v) r(s) \right]\Theta (s-4m^2)\, ,
\end{equation}
where
\vspace{-2mm}
\begin{eqnarray}\label{vTg}
T(v)=v\frac{3-v^2}{2}\, ,\> \>
g(v)=\frac{4\pi}{3}\left[\frac{\pi}{2v}-\frac{3+v}{4}
\left(\frac{\pi}{2}-\frac{3}{4\pi} \right) \right]\, , \> \>
v=\sqrt{1-\frac{4m^2}{s}}\, .
\end{eqnarray}

We introduce effective quark masses, which accumulate some non-perturbative
contributions and turn out to be close to the constituent masses. In the
description of the non-strange vector component of the $D$-function we
neglect the difference of the quark mass values and set $m_u=m_d=m$.

The threshold resummation factor (\ref{S-factor-relativistic}) leads to the
following modification of the expression (\ref{R_approx})
\begin{equation}\label{R-approx1}
{\cal{R}}_V(s)=T(v)\,\left[S(\chi)-\frac{1}{2}X(\chi)+g(v)\,r(s)
\right]\Theta (s-4m^2)\,,
\end{equation}
which one can use to calculate the vector component of the $R_\tau$ ratio
\begin{equation}\label{R_tau_V}
R_{\tau,V}=3 S_{\rm EW}|V_{ud}|^2\,\int_0^{M_{\tau}^2}\frac{ds}{M_{\tau}^2}
\left(1-\frac{s}{M_{\tau}^2} \right)^2 \left(1+\frac{2s}{M_{\tau}^2}\right)
{\cal R}_V(s)\, ,
\end{equation}
where $S_{\rm EW}$ denotes the electroweak factor~\cite{S_EW} and
the light vector $D$-function is defined by Eq.~(\ref{D_R}) with
${\cal R}\to{\cal R}_V$.

\subsection{Results}
\begin{figure}[thb]
\centerline{\epsfig{file=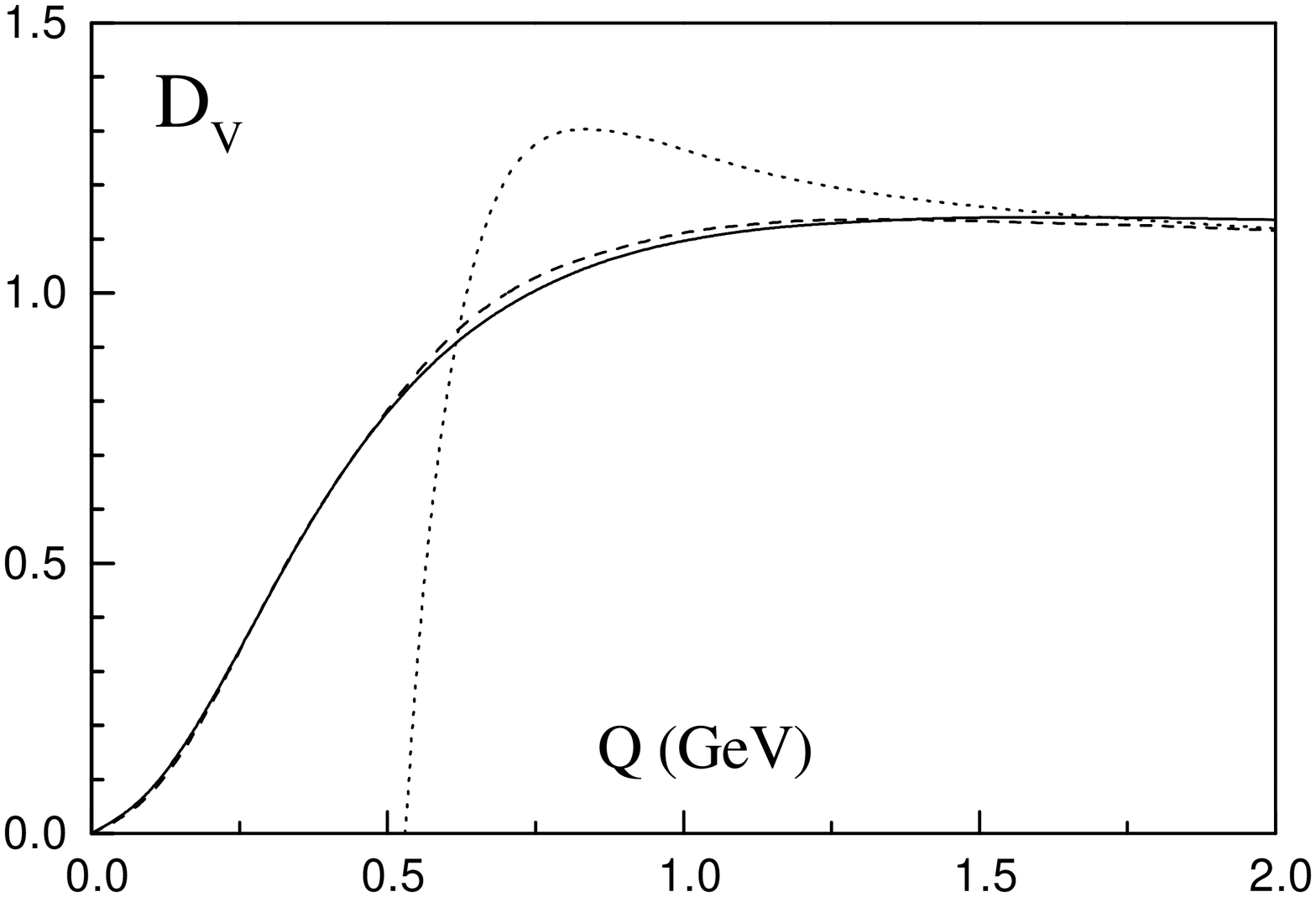,width=11.0cm}}
\caption{The light Adler function corresponding to the non-strange vector
channel of $\tau$ decay data. The solid curve is the APT result. The
experimental curve (dashed line) corresponding to the ALEPH data and the
perturbative result with power corrections (dotted line) are taken from
Ref.~\protect\cite{PPRaf}.}
\label{d-v}
\end{figure}

We derive the $D_V$-function and the value of $R_{\tau,V}$ by using
Eq.~(\ref{R-approx1}) with the expression (\ref{r3}) for $r(s)$ and the
timelike running coupling (\ref{time3}) in the $S$-factor. In Fig.~\ref{d-v}
we plot this $D$-function (solid curve) which was calculated by using the
parameter $\Lambda=420\,{\rm{MeV}}$ and the value of the quark masses
$m_u=m_d=250\,{\rm{MeV}}$. Note that, practically, the same values of the
light quark masses have been obtained in Refs.~\cite{20,21}. These values are
close to the constituent quark masses and incorporate some nonperturbative
effects. The shape of the infrared tail of the $D$-function is rather
sensitive to the value of these masses. We obtain the value of
$R_{\tau,V}=1.77$ which agrees well with the experimental data presented by
the ALEPH, $R_{\tau,V}^{\rm{expt}}=1.775\pm0.017$ \cite{ALEPH98}, and the
OPAL, $R_{\tau,V}^{\rm{expt}}=1.764\pm0.016$ \cite{OPAL99}, collaborations.
The value of the parameter $\Lambda$ obtained here is slightly more than in
the standard analysis. For example, the ALEPH Collaboration result is
$\Lambda=370\pm13_{\rm{expt}}\pm38_{\rm{theor}}$~MeV. In this
connection, note that a more accurate account of the so-called $\pi^2$-terms
in the treatment of high energy experimental data for timelike processes
leads to an increase in the value of $\alpha_S$ \cite{Shirkov-time}. It should
be emphasized that the value of $\Lambda$ turns out to be significantly
smaller than the value extracted from the $\tau$ data in Ref.~\cite{tau00},
which
shows the importance of threshold effects. [Recall that in Fig.~\ref{d-v} 
we also show
the experimental curve (dashed line) and the curve corresponding to the
perturbative result with power corrections (dotted line) which are taken
from Ref.~\cite{PPRaf}.]
Note also that using the above values of quark masses,
we find that the one-loop APT description gives a result
(not shown in the Figure) which is close to the three-loop one and
reproduces well the experimental behavior of the $D$-function.

\section{Conclusion}
In this paper we have considered the Adler function corresponding to
the non-strange vector channel data from $\tau$ decay. This function, defined
in the Euclidean region, is a smooth function and represents a convenient
testing ground for theoretical methods. We have
proposed the method of the ``light'' vector $D$-function description. The
conventional method of approximating this function as a sum of perturbative
terms and power corrections cannot describe the low energy scale region
because both the logarithmic and power expansions diverge at small
momenta. We have used the analytic approach to QCD which is not in conflict
with the general principles of the theory and, in the infrared region, has a
regular behavior. The new relativistic resummation factor, which corresponds
to a QCD-like quark-antiquark quasipotential, has been used to incorporate
the threshold effects. We have shown that our approach allows us to describe
well the experimental data for $\tau$-lepton decay in terms of the
$D$-function
down to the lowest energy scale and for $R_\tau$ in the non-strange vector
channel. We have found that the influence of relativistic threshold
resummation is important and leads to an significant reduction of the value
of the QCD scale parameter $\Lambda$ extracted from the $\tau$ data.

\section*{Acknowledgments}
The authors would like to thank D.V.~Shirkov and A.N.~Sissakian for interest
in this work. Partial support of the work by the US National Science
Foundation, grant PHY-9600421, and by the US Department of Energy, grant
DE-FG-03-98ER41066, and by the RFBR, grants 99-01-00091, 99-02-17727,
00-15-96691 is gratefully acknowledged. The work of ILS and OPS is also
supported in part by the University of Oklahoma, through its College of Arts
and Science, the Vice President for Research, and the Department of Physics
and Astronomy.

\end{document}